\documentclass[aps,prb,twocolumn,groupedaddress,showpacs,showkeys]{revtex4}

\usepackage{graphicx}
\usepackage{bm}
\usepackage{amsmath}

\begin{document}


\title{
Crossover from impurity-induced ordered phase to uniform antiferromagnetic phase under hydrostatic pressure in the doped spin-gap system TlCu$_{1-x}$Mg$_x$Cl$_3$
}


\author{Hideyo Imamura}
\author{Toshio Ono}
\email[]{o-toshio@lee.phys.titech.ac.jp}
\author{Kenji Goto}
\author{Hidekazu Tanaka}
\affiliation{Department of Physics, Tokyo Institute of Technology, Oh-okayama, Meguro-ku, Tokyo 152-8551, Japan}


\date{\today}

\begin{abstract}
Magnetic phase transition under hydrostatic pressure in TlCu$_{0.988}$Mg$_{0.012}$Cl$_3$ was investigated by magnetization measurements. The parent compound TlCuCl$_3$ is a coupled spin dimer system, which undergoes a pressure-induced quantum phase transition from a gapped ground state to an antiferromagnetic state at $P_{\rm c} = 0.42$ kbar due to the shrinkage of the gap. At ambient pressure, the present doped system exhibits impurity-induced magnetic ordering at $T_{\rm N}=2.5$ K. With increasing pressure, $T_{\rm N}$ increases. This is because the effective exchange interaction $J_{\rm eff}$ between unpaired spins is enhanced by the shrinkage of the gap. With a further increase in pressure, the present system undergoes a phase transition to a uniform antiferromagnetic phase due to the closing of the triplet gap in the intact dimers. The crossover from the impurity-induced ordered phase to the uniform antiferromagnetic phase occurs at $P\simeq 1.3$ kbar.
\end{abstract}


\maketitle

\section{Introduction}
Singlet ground states with excitation gaps (spin gaps) have been observed in many quantum spin systems. Most gapped ground states are composed of singlet spin dimers. The excitations from the ground state are triplet excitations, which are regarded as bosons with a hard core repulsion. Recently, magnetic phase transitions in the gapped spin systems induced by external magnetic field, pressure and impurities have been attracting considerable attention. In these phase transitions, the triplet excitations play important roles. Field- and pressure-induced magnetic orderings are caused by the softening or the Bose-Einstein condensation of the triplet modes \cite{Giamarchi,Nikuni,Rice,Rueegg,Matsumoto,Kawashima,Misguich,Nohadani}. 

When a small number of nonmagnetic ions are substituted for magnetic ions, singlet spin dimers are partially broken, so that unpaired spins are produced near the nonmagnetic ions. Unpaired spins can interact through the effective exchange interaction that is mediated by the triplet excitations in intact dimers \cite{Sigrist,Imada,Yasuda,Mikeska}. This mechanism is analogous to that for the RKKY interaction between magnetic impurities in a nonmagnetic metal. The effective exchange interaction $J_{\rm eff}$ depends on the spin gap $\varDelta$ in intact dimers and $J_{\rm eff}$ increases with decreasing $\varDelta$. This effective exchange interaction can cause the ordering of unpaired spins. As a result, the small staggered magnetic order appears in intact dimers. Such impurity-induced magnetic ordering is observed in many gapped spin systems doped with nonmagnetic impurities \cite{Hase,Masuda1,Azuma,Uchiyama,Masuda2}.

This study is concerned with impurity- and pressure-induced magnetic orderings in TlCu$_{1-x}$Mg$_x$Cl$_3$. The parent compound TlCuCl$_3$ is an $S=1/2$ coupled spin dimer system with an antiferromagnetic intradimer exchange interaction $J/k_{\rm B} = 65.9$ K \cite{Takatsu,Cavadini,Oosawa1}. The magnetic ground state is a spin singlet with an excitation gap $\varDelta/k_{\rm B}$ = 7.5 K \cite{Takatsu,Shiramura,Oosawa2}. The small gap compared with the intradimer exchange interaction $J$ is attributed to strong interdimer exchange interactions \cite{Cavadini,Oosawa1}. 

At ambient pressure, TlCuCl$_3$ is paramagnetic down to zero temperature because of the spin gap. However, TlCuCl$_3$ undergoes magnetic ordering under hydrostatic pressure \cite{Goto}. Neutron scattering experiments \cite{Oosawa3,Oosawa4,Rueegg2} demonstrate that the lowest excitation gap at $\boldsymbol{Q}=(0,0,1)$ closes by the application of hydrostatic pressure, so that antiferromagnetic ordering characterized by the same ordering vector occurs. These results indicate that the pressure-induced magnetic ordering in TlCuCl$_3$ is a quantum phase transition due to the closing of the spin gap. The critical pressures obtained through magnetization measurement and a neutron scattering experiment are $P_{\rm c}=0.42$ kbar \cite{Goto} and 1.07 kbar \cite{Rueegg2}, respectively. There is a discrepancy between the critical pressures obtained through the macroscopic and microscopic measurements, and the cause of the discrepancy is not clear.

Magnetization measurement and a neutron scattering experiment revealed that doped TlCu$_{1-x}$Mg$_x$Cl$_3$ with $x\leq 0.03$ exhibits impurity-induced magnetic ordering \cite{Oosawa5,Oosawa6}. The ordering vector of the impurity-induced phase $\boldsymbol{Q}=(0,0,1)$ is the same as those of the field- and pressure-induced ordered phases in the parent compound TlCuCl$_3$ \cite{Tanaka,Oosawa3}. The easy axis lies in the $(0,1,0)$ plane and is close to the $[2,0,1]$ direction \cite{Oosawa5}. Spin flop transition was observed at $H_{\rm sf} \simeq 0.35$ T, which is almost independent of $x$. Triplet excitations with a finite gap for intact dimers were also observed \cite{Oosawa6}. The gap increases below the transition temperature $T_{\rm N}$, which can be attributed to the evolution of the staggered order in intact dimers \cite{Mikeska}. It is noted that the gap for TlCu$_{1-x}$Mg$_x$Cl$_3$ is larger than that for a pure system even above $T_{\rm N}$. 

As mentioned above, the effective interaction $J_{\rm eff}$ between unpaired spins is enhanced as the gap decreases. This leads to an increase of $T_{\rm N}$ in the impurity-induced magnetic ordering. Although there are many experimental studies of impurity-induced magnetic ordering in doped spin gap systems, the systematic study of impurity-induced magnetic ordering by regulating the gap $\varDelta$ is limited. Because the spin gap in TlCuCl$_3$ shrinks under hydrostatic pressure, we expect an increase in $T_{\rm N}$ with pressure in TlCu$_{1-x}$Mg$_x$Cl$_3$. To investigate the systematic change in the magnetic ordering in TlCu$_{1-x}$Mg$_x$Cl$_3$, we carried out magnetization measurements under hydrostatic pressure.

\section{Experimental}
Doped TlCu$_{1-x}$Mg$_x$Cl$_3$ crystals were prepared as follows: we first prepared single crystals of TlCuCl$_3$ and TlMgCl$_3$. Mixing TlCuCl$_3$ and TlMgCl$_3$ in a ratio of $(1-x) : x$, we prepared TlCu$_{1-x}$Mg$_x$Cl$_3$ by the vertical Bridgman method. Single crystals of $\sim 0.5$ cm$^3$ were obtained. The magnesium concentration $x$ was analyzed by inductively coupled plasma$-$optical emission spectroscopy (ICP$-$OES). In the present study, we use a sample with $x=0.012$. 

The magnetizations were measured at temperatures down to 1.8 K under magnetic fields of up to 7 T using a SQUID magnetometer (Quantum Design MPMS XL). Pressures of up to 6.21 kbar were applied using a cylindrical high-pressure clamp cell designed for use with the SQUID magnetometer \cite{Uwatoko}. A sample of size $2.5\times 2.5\times 5$ mm$^3$ was set in the cell with its $[2, 0, 1]$ direction parallel to the cylindrical axis. The $[2,0,1]$ direction is parallel to two cleavage planes, $(0,1,0)$ and $(1,0,\bar{2})$. A magnetic field was applied along the $[2,0,1]$ direction. As pressure-transmitting fluid, a mixture of Fluorinert FC70 and FC77 was used. The pressure was calibrated with the superconducting transition temperature $T_{\rm c}$ of tin placed in the pressure cell. The diamagnetism of tin was measured at $H=10$ and 50 Oe to determine $T_{\rm c}$ after removing the residual magnetic flux trapped in the superconducting magnet. The accuracy of the pressure was 0.1 kbar for the absolute value and 0.05 kbar for the relative value.

\section{Results and Discussion}
\begin{figure}[htbp]
  \begin{center}
    \includegraphics[keepaspectratio=true,width=80mm]{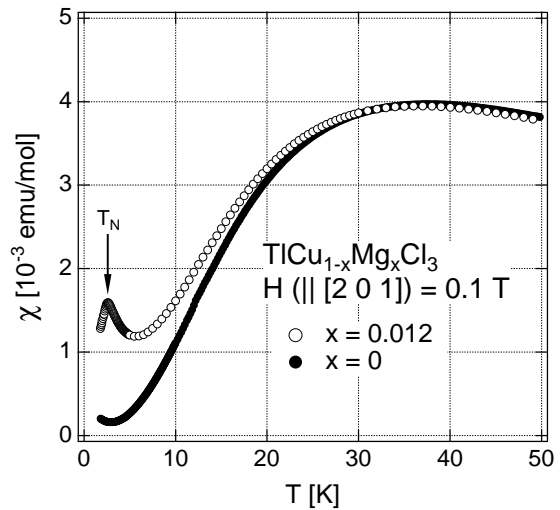}
  \end{center}
  \caption{Temperature dependences of magnetic susceptibilities in TlCuCl$_3$ and TlCu$_{0.988}$Mg$_{0.012}$Cl$_3$ measured at ambient pressure for $H\parallel [2,0,1]$. The arrow denotes N\'{e}el temperature $T_{\rm N}$.}
  \label{fig:chi_0.eps}
\end{figure}

Figure \ref{fig:chi_0.eps} shows the temperature dependences of the magnetic susceptibilities $\chi=M/H$ of pure TlCuCl$_3$ and TlCu$_{0.988}$Mg$_{0.012}$Cl$_3$ under ambient pressure. A magnetic field of $H=0.1$ T was applied along the $[2,0,1]$ direction. With decreasing temperature, the magnetic susceptibility in TlCu$_{0.988}$Mg$_{0.012}$Cl$_3$ displays a broad maximum at $T\simeq 36$ K and then a decrease, as observed in pure TlCuCl$_3$ \cite{Takatsu,Oosawa2}. However, the magnetic susceptibility in the present doped system increases again below 5 K, and then exhibits a sharp cusplike anomaly at $T_{\rm N}=2.5$ K, which is indicative of magnetic ordering. The decrease in $\chi$ below 36 K arises from the formation of singlet states in intact dimers. The increase in $\chi$ below 5 K is attributed to the paramagnetic susceptibility of unpaired spins and their ordering gives rise to the decrease in $\chi$ below $T_{\rm N}=2.5$ K. From the previous magnetization measurements and neutron scattering experiments on TlCu$_{1-x}$Mg$_x$Cl$_3$ \cite{Oosawa5,Oosawa6}, it was generally found that the magnetic structure of the impurity-induced ordered phase is described by the two-sublattice model, and that the easy-axis lies in the $(0,1,0)$ plane and is close to the the $[2,0,1]$ direction; e.g., the angle between the easy-axis and the $[2,0,1]$ direction is 13$^\circ$ for $x=0.022$ \cite{Oosawa5}. Thus, the susceptibility for $H\parallel [2,0,1]$ approximates to the parallel susceptibility ${\chi}_{\parallel}$ that decreases toward zero with decreasing temperature. For this reason, the susceptibility for $H\parallel [2,0,1]$ decreases below $T_{\rm N}$.

\begin{figure}[htbp]
  \begin{center}
    \includegraphics[keepaspectratio=true,width=80mm]{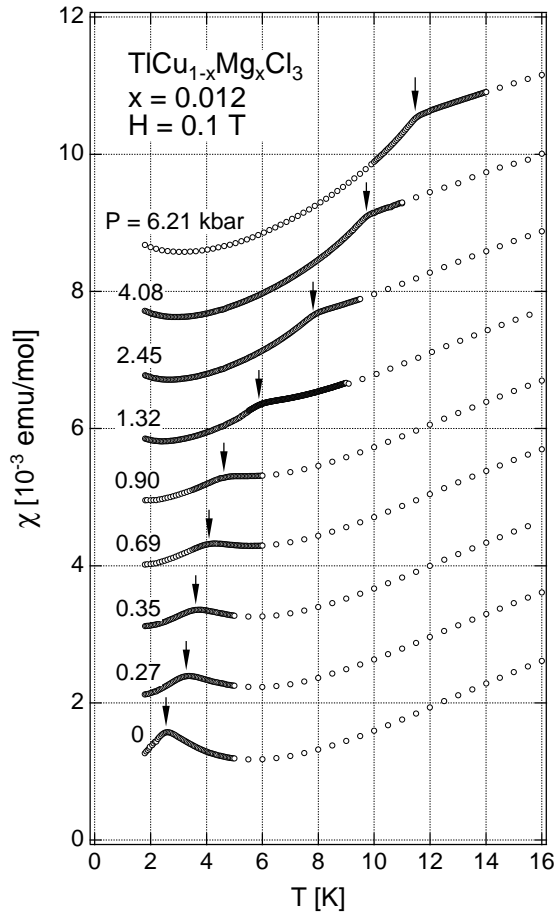}
  \end{center}
  \caption{Low-temperature magnetic susceptibilities in TlCu$_{0.988}$Mg$_{0.012}$Cl$_3$ measured at various pressures for $H\parallel [2,0,1]$. For clarity, the values of the susceptibilities are shifted upward consecutively by $1\times 10^{-3}$ emu/mol with increasing pressure. Arrows denote N\'{e}el temperatures $T_{\rm N}$.}
  \label{fig:chi_low.eps}
\end{figure}

\begin{figure}[htbp]
  \begin{center}
    \includegraphics[keepaspectratio=true,width=80mm]{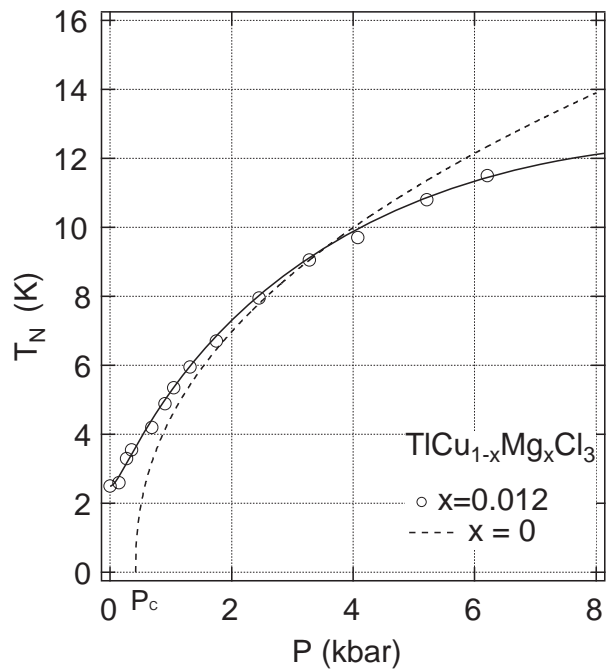}
  \end{center}
  \caption{Transition temperature $T_{\rm N}$ as function of pressure in TlCu$_{0.988}$Mg$_{0.012}$Cl$_3$. The solid line is a visual guide. The dashed line denotes the transition temperature of pressure-induced magnetic ordering in pure TlCuCl$_3$ \cite{Goto}.}
  \label{fig:TN.eps}
\end{figure}

Figure \ref{fig:chi_low.eps} shows the low-temperature susceptibilities in TlCu$_{0.988}$Mg$_{0.012}$Cl$_3$ measured at various pressures. 
Arrows in Fig. \ref{fig:chi_low.eps} denote the transition temperatures.  The pressure dependence of $T_{\rm N}$ is plotted in Fig. \ref{fig:TN.eps}. The transition temperature $T_{\rm N}$ of impurity-induced magnetic ordering increases monotonically with pressure. The impurity-induced magnetic ordering is caused by the effective interaction $J_{\rm eff}$ between unpaired spins. This effective interaction is determined by the distance $r$ between unpaired spins and the spin correlation length ${\xi}$, which is inversely proportional to the spin gap $\varDelta$. If ${\xi}$ is small, $J_{\rm eff}\propto \exp (-r/{\xi})$ \cite{Sigrist,Imada,Mikeska}. The effective interaction $J_{\rm eff}$ increases as the gap decreases. In the present doped system, the gap shrinks with increasing pressure. Thus, the $J_{\rm eff}$ of the present doped system increases with increasing pressure. This gives rise to the increase in $T_{\rm N}$ for impurity-induced magnetic ordering.

In Fig. \ref{fig:TN.eps}, we appended the transition temperature $T_{\rm N}$ of pressure-induced magnetic ordering in pure TlCuCl$_3$ (dashed line). The pressure dependence of $T_{\rm N}$ is described by the power law $T_{\rm N} = 5.66(P-P_{\rm c})^{\beta}$ K with $\beta=0.44$ and the critical pressure $P_{\rm c}=0.42$ kbar. This pressure-induced magnetic ordering is caused by the shrinkage of the triplet gap under hydrostatic pressure \cite{Matsumoto,Goto,Oosawa3,Oosawa4,Rueegg2}. The gap shrinks either when the intradimer interaction is reduced or when the interdimer interaction is enhanced by an applied pressure. In KCuCl$_3$, applied pressure produces a decrease in intradimer interaction and an increase in interdimer interaction \cite{Goto2}. These pressure dependences of the intradimer and interdimer interactions both act synergistically to reduce the gap. We expect that the same is true of TlCuCl$_3$.
The ground state of the pressure-induced ordered phase is represented by a coherent superposition of the singlet and the triplet components on each dimer. With increasing pressure in TlCuCl$_3$, the contribution of the triplet components to the ground state is enhanced, which leads to an increase in $T_{\rm N}$.

With increasing pressure, the height of the susceptibility cusp at $T_{\rm N}$ is suppressed and the cusplike anomaly changes to a bend anomaly at $P_0\simeq 1.3$ kbar. The bend anomaly at $T_{\rm N}$ is characteristic of pressure-induced magnetic transition to the uniform antiferromagnetic phase, as observed in pure TlCuCl$_3$ \cite{Goto}. In the uniform antiferromagnetic phase, the triplet excitation gap in intact dimers closes completely. As mentioned above, the cusplike anomaly in susceptibility is characteristic of impurity-induced magnetic phase transition. Since the changes in the susceptibility anomaly and $T_{\rm N}$ at $P_0$ are not rapid but gradual, we can deduce that the impurity- and pressure-induced uniformly ordered phases are contiguous and that the crossover between these two phases occurs at $P_0\simeq 1.3$ kbar. This crossover pressure is larger than the critical pressure $P_{\rm c}$ of the pressure-induced magnetic ordering in pure TlCuCl$_3$, where $P_{\rm c}$ obtained by magnetization measurements and neutron inelastic scattering is 0.42 kbar \cite{Goto} and 1.07 kbar \cite{Rueegg2}, respectively. In TlCuCl$_3$, the triplet excitation gap is $\varDelta/k_{\rm B} = 7.5$ K \cite{Takatsu,Shiramura,Oosawa2}. The triplet gap remains even in the doped TlCu$_{1-x}$Mg$_x$Cl$_3$ system and is larger than $\varDelta$ in TlCuCl$_3$ \cite{Oosawa6}. It is considered that when nonmagnetic ions are doped, the hopping of the spin triplet to the doped dimer site is forbidden, so that the hopping amplitude is generally suppressed. This gives rise to the suppression of the dispersion range of the spin triplets and leads to an enhancement in the lowest excitation energy. For $x\approx 0.03$, $\varDelta/k_{\rm B} \simeq 15$ K \cite{Oosawa6}. Thus, the triplet gap in TlCu$_{0.988}$Mg$_{0.012}$Cl$_3$ is expected to be $\varDelta/k_{\rm B} \simeq 10$ K. We infer that the increase in the triplet gap with doping is responsible for the increase in the crossover pressure $P_0$, against $P_{\rm c}=0.42$ kbar for TlCuCl$_3$.

As shown in Fig. \ref{fig:TN.eps}, the phase boundaries for doped and pure systems intersect at $P\simeq 3$ kbar. For $P > 3$ kbar, $T_{\rm N}$ for TlCu$_{0.988}$Mg$_{0.012}$Cl$_3$ is lower that $T_{\rm N}$ for TlCuCl$_3$. The separation of phase boundaries for pure and doped systems is enlarged with increasing pressure. If this difference in $T_{\rm N}$ is due to the difference in the triplet gap $\varDelta$, both phase boundaries should be parallel. Thus, the suppression of $T_{\rm N}$ for pressure-induced magnetic ordering in the present doped system should be ascribed to the disorder produced by doping, because the disorder acts to prevent triplet components from forming a coherent state in intact dimers. 

\begin{figure}[htbp]
  \begin{center}
    \includegraphics[keepaspectratio=true,width=80mm]{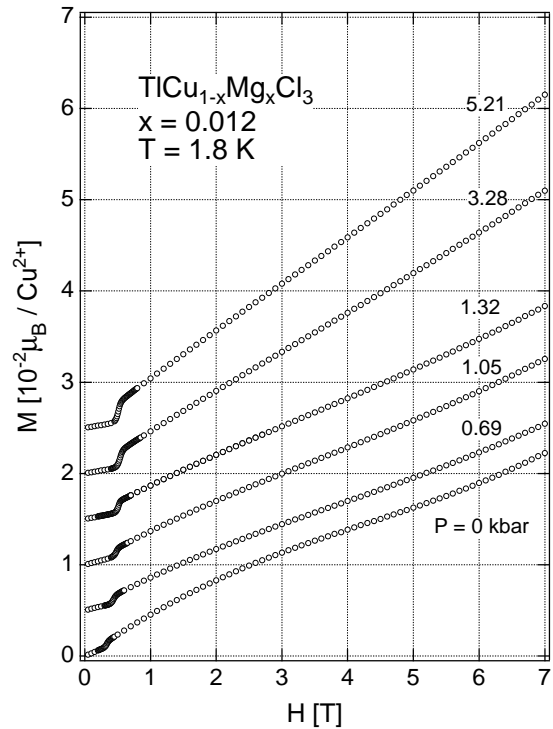}
  \end{center}
  \caption{Magnetization curves in TlCu$_{0.988}$Mg$_{0.012}$Cl$_3$ measured at $T=1.8$ K under various pressures. The values of magnetizations are shifted upward consecutively by $5\times 10^{-3}$ $\mu_{\rm B}$/Cu$^{2+}$ with increasing pressure.}
  \label{fig:M-H.eps}
\end{figure}

Figure \ref{fig:M-H.eps} shows magnetization curves for $H\parallel [2,0,1]$ in TlCu$_{0.988}$Mg$_{0.012}$Cl$_3$ measured at $T=1.8$ K at various pressures. Magnetization increases with a finite slope, and exhibits a jump indicative of spin-flop transition at $H_{\rm sf} = 0.34\sim 0.54$ T with increasing magnetic field. For $P\leq 1.05$ kbar, the magnetization curve for $H > H_{\rm sf}$ displays a concave field dependence for $H < 5$ T, which may be explained by the magnetization process of unpaired spins. For $P > 1.32$ kbar, magnetization increases linearly, as observed in the pressure-induced uniformly ordered state in pure TlCuCl$_3$ \cite{Goto}. In this pressure region, the magnetization behavior is mainly dominated by the triplet components in intact dimers. 

\begin{figure}[htbp]
  \begin{center}
    \includegraphics[keepaspectratio=true,width=80mm]{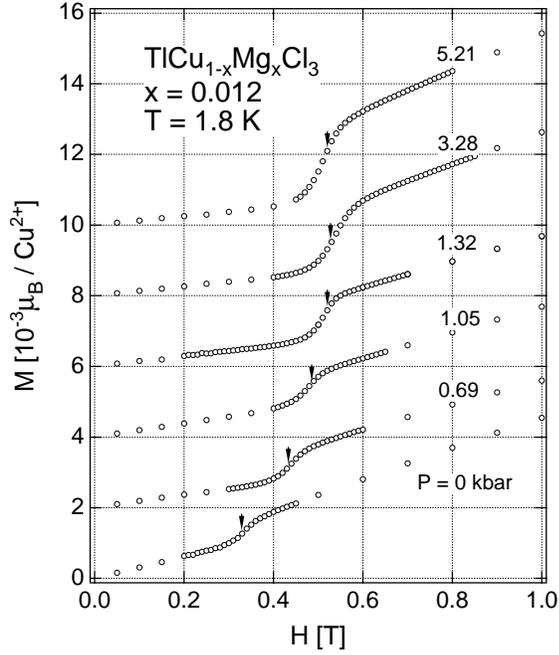}
  \end{center}
  \caption{Magnetization curves around spin-flop field region measured at $T=1.8$ K under various pressures. The values of magnetizations are shifted upward consecutively by $2\times 10^{-3}$ $\mu_{\rm B}$/Cu$^{2+}$ with increasing pressure.}
  \label{fig:M-H_low.eps}
\end{figure}

\begin{figure}[htbp]
  \begin{center}
    \includegraphics[keepaspectratio=true,width=80mm]{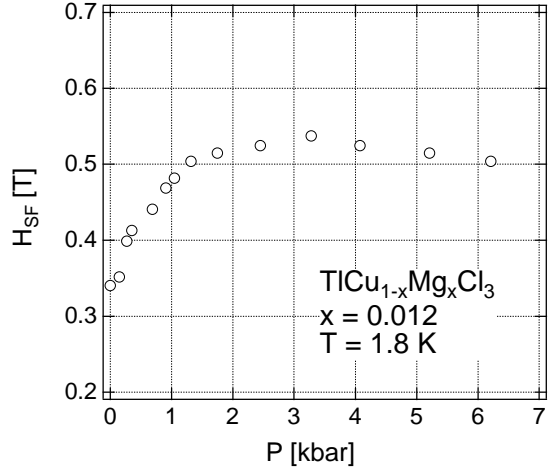}
  \end{center}
  \caption{Spin-flop field $H_{\rm sf}$ at $T=1.8$ K as function of pressure in TlCu$_{0.988}$Mg$_{0.012}$Cl$_3$.}
  \label{fig:Hsf.eps}
\end{figure}

Figure \ref{fig:M-H_low.eps} shows the enlargement of magnetization curves around the spin-flop field region. Fields for the spin-flop transition are indicated by arrows. We assign the field at which ${\rm d}M/{\rm d}H$ exhibits the peak anomaly as the spin-flop $H_{\rm sf}$. The pressure dependence of $H_{\rm sf}$ is plotted in Fig. \ref{fig:Hsf.eps}. With increasing pressure, the spin-flop transition field $H_{\rm sf}$ increases from 0.34 T and reaches 0.50 T at the crossover pressure $P_0\simeq 1.3$ kbar. Although the magnitude of the ordered moment increases with applied pressure, $H_{\rm sf}$ for $P > P_0$ has little dependence on pressure, as observed in the pressure-induced ordered state in pure TlCuCl$_3$ \cite{Goto}, for which $H_{\rm sf}\simeq 0.7$ T. The reason that $H_{\rm sf}$ is almost independent of pressure in the pressure-induced ordered state is given in ref. \onlinecite{Goto}. 
At present, we have no clear explanation for the increase in $H_{\rm sf}$ for $P < P_0$. From the previous magnetization measurements on TlCu$_{1-x}$Mg$_x$Cl$_3$, it was found that $H_{\rm sf}$ is almost independent of $x$, although the size of the magnetization jump at $H_{\rm sf}$ increases with $x$ \cite{Oosawa5}. Therefore, we infer that the local interactions between the unpaired spins and paired spins in the intact dimers determine $H_{\rm sf}$ for $P > P_0$.

\begin{figure}[htbp]
  \begin{center}
    \includegraphics[keepaspectratio=true,width=80mm]{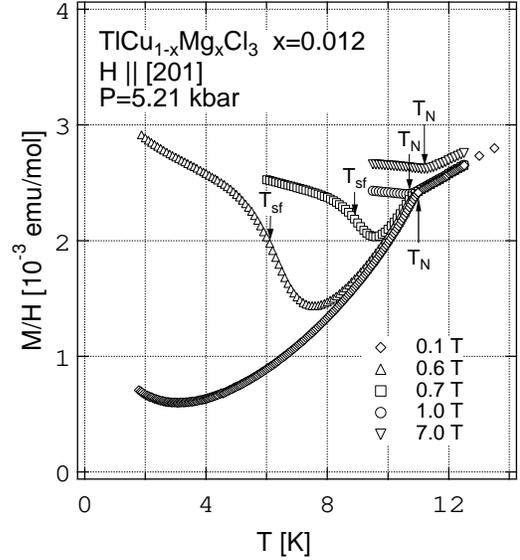}
  \end{center}
  \caption{Magnetizations in TlCu$_{0.988}$Mg$_{0.012}$Cl$_3$ under $P=5.21$ kbar as functions of temperature measured at various magnetic fields parallel to $[2,0,1]$ direction. Arrows denote transition temperatures $T_{\rm N}$ and $T_{\rm sf}$.}
  \label{fig:M-T5.21.eps}
\end{figure}

\begin{figure}[htbp]
  \begin{center}
    \includegraphics[keepaspectratio=true,width=80mm]{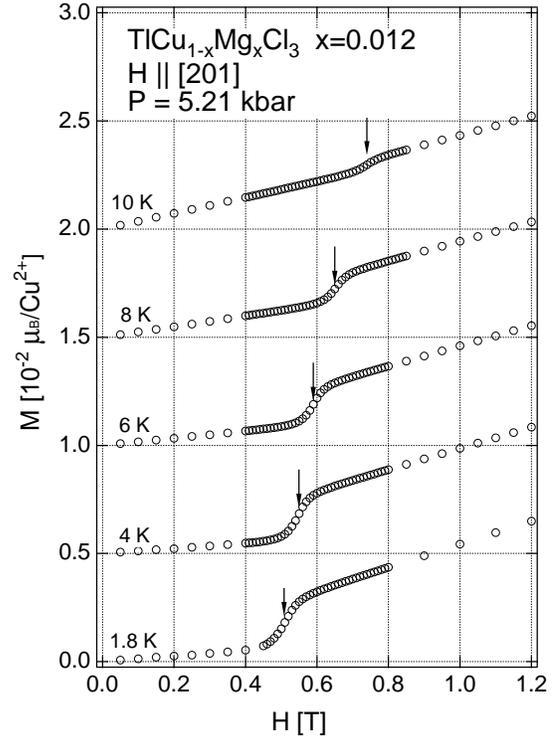}
  \end{center}
  \caption{Low-field magnetization curves in TlCu$_{0.988}$Mg$_{0.012}$Cl$_3$ measured at various temperatures for $H\parallel [2,0,1]$ at $P=5.21$ kbar. Arrows denote spin-flop fields $H_{\rm sf}$.}
  \label{fig:M-H5.21.eps}
\end{figure}

\begin{figure}[htbp]
  \begin{center}
    \includegraphics[keepaspectratio=true,width=80mm]{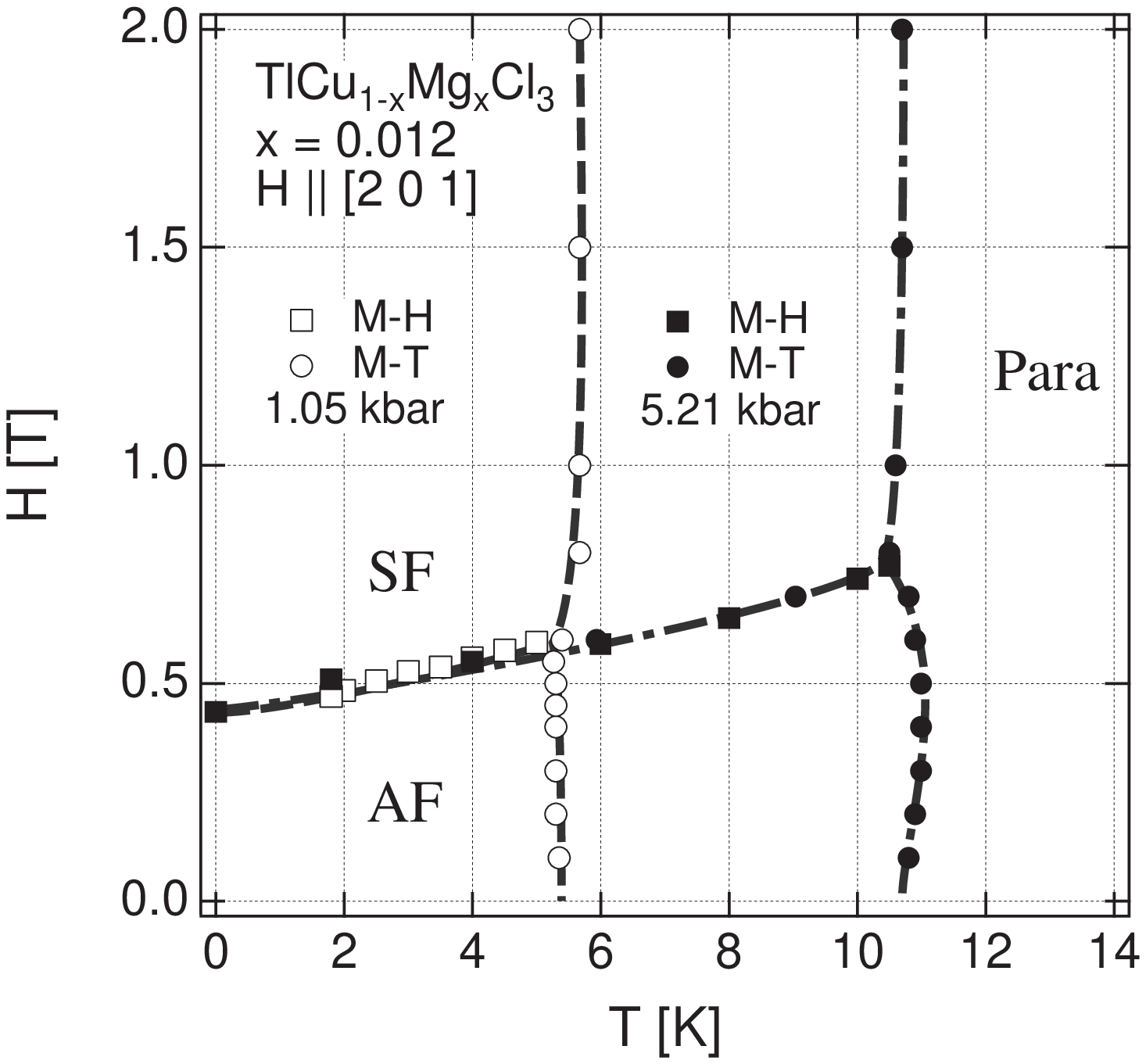}
  \end{center}
  \caption{Magnetic phase diagrams for $H\parallel [2,0,1]$ in TlCu$_{0.988}$Mg$_{0.012}$Cl$_3$ obtained at $P=5.21$ (closed marks) and 1.05 kbar (open marks).}
  \label{fig:Phase.eps}
\end{figure}

To obtain a magnetic phase diagram for magnetic field vs temperature in TlCu$_{0.988}$Mg$_{0.012}$Cl$_3$, we measured the temperature and the field scans of magnetization at $P=1.05$ kbar ($ < P_0$) and 5.21 kbar ($ \gg P_0$). The results for $P=5.21$ kbar are shown in Figs. \ref{fig:M-T5.21.eps} and \ref{fig:M-H5.21.eps}. In Fig. \ref{fig:M-T5.21.eps}, the transition temperature for the transition from the paramagnetic phase to the antiferromagnetic phase or spin-flop phase is denoted as $T_{\rm N}$ and that for the transition from the antiferromagnetic phase to the spin-flop phase is denoted as $T_{\rm sf}$. The spin-flop field $H_{\rm sf}$ increases with temperature. The transition data for $P=1.05$ and 5.21 kbar are plotted in Fig. \ref{fig:Phase.eps}. The phase diagrams for these pressures are similar to those for conventional antiferromagnets with easy-axis anisotropy. 

\begin{figure}[htbp]
  \begin{center}
    \includegraphics[keepaspectratio=true,height=60mm]{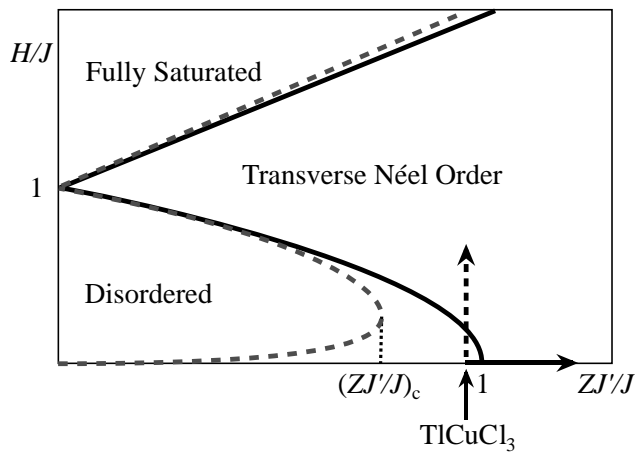}
  \end{center}
  \caption{Phase diagram of ground state proposed by Mikeska \textit{et al}\cite{Mikeska}. Solid and dashed lines are boundaries for the pure and doped systems, respectively . The thin arrow denotes the ($ZJ'/J$) corresponding to TlCuCl$_3$. The thick solid and dashed arrows denote the changes of the ground state induced by applications of pressure and magnetic field, respectively.}
  \label{fig:mikeska.eps}
\end{figure}

Mikeska {\it et al.} \cite{Mikeska} have theoretically investigated the effect of doping nonmagnetic impurities on the ground state in the external field in the coupled $S=1/2$ Heisenberg spin dimer system. They presented the schematic phase diagram in the $(ZJ'/J, H/J)$ plane, as shown in Fig. \ref{fig:mikeska.eps}, where $J$ and $J'$ are the intradimer and interdimer interactions, respectively, and $Z$ is the coordination number. Solid and dashed lines are phase boundaries for the pure and doped systems, respectively. Their results are summarized as follows: 
impurity-induced antiferromagnetic ordering occurs at a zero field for finite impurity concentration $x$. When $(ZJ'/J)$ is small, there are three critical fields $H_{\rm{c1}}$, $H_{\rm{c2}}$ and $H_{\rm s}$. $H_{\rm s}$ is the saturation field at which all spins are fully polarized. For $H < H_{\rm{c1}}$, the unpaired spins form an antiferromagnetic order through the effective exchange interaction $J_{\rm eff}$, and a small transverse staggered order also occurs in intact dimers. Unpaired spins are fully polarized at $H = H_{\rm{c1}}$, and the small transverse staggered order in intact dimers vanishes. The critical field $H_{\rm{c1}}$ is proportional to the impurity concentration $x$. At the second critical field $H_{\rm{c2}}$, the triplet gap for intact dimers closes and the transverse staggered ordering occurs again in intact dimers. 
The ground state between $H_{\rm{c1}}$ and $H_{\rm{c2}}$ is disordered. With increasing $(ZJ'/J)$, the field range of the disordered state shrinks. The disordered state vanishes at $(ZJ'/J)_{\rm c}$, above which impurity- and field-induced ordered states merge. The point given by $(ZJ'/J) = 1$ for the pure system is the quantum critical point that separates the gapped disordered state and the antiferromagnetic ordered state. For the pure system ($x=0$), $(ZJ'/J)_{\rm c}=1$. With increasing $x$, $(ZJ'/J)_{\rm c}$ decreases.

Because the critical field corresponding to the gap and the saturation field for TlCuCl$_3$ are $(g/2)H_{\rm c}$ = 5.5 T and $(g/2)H_{\rm s} \sim$ 90 T, respectively, the exchange parameter should be located at the point indicated by the thin arrow in Fig. \ref{fig:mikeska.eps}. The $(ZJ'/J)$ corresponding to TlCuCl$_3$ is close to unity and larger than the $(ZJ'/J)_{\rm c}$ for $x = 0.012$. This was confirmed by Fujisawa {\it et al.} \cite{Fujisawa} through specific heat and neutron scattering experiments in magnetic fields on TlCu$_{1-x}$Mg$_{x}$Cl$_3$ with $x = 0.0088$ and 0.015. Their results revealed that the impurity- and field-induced ordered phases are contiguous; i.e., these two phases are the same. The experiments by Fujisawa {\it et al.} correspond to the scan along the thick dashed arrow in Fig. \ref{fig:mikeska.eps}. Since the excitation gap in TlCuCl$_3$ decreases under hydrostatic pressure, the application of pressure corresponds to the scan along the thick solid arrow. Therefore, no phase transition is expected to occur with increasing $(ZJ'/J)$ in TlCu$_{1-x}$Mg$_{x}$Cl$_3$ with $x = 0.012$. This is in accordance with our result that the impurity- and pressure-induced uniformly ordered phases in the present doped system are contiguous. The effects of application of pressure and magnetic field are analogous in decreasing the triplet gap in intact dimers.  
The present study demonstrated that the gapped ground state observed in pure TlCuCl$_3$ is completely wiped out by the small amount of doping, and that the quantum phase transition at $(ZJ'/J) = 1$ is smeared to become a crossover.

\section{Conclusion}
We have presented the results of magnetization measurements on the doped spin-gap system TlCu$_{1-x}$Mg$_{x}$Cl$_3$ with $x = 0.012$ under hydrostatic pressure. At ambient pressure, the present doped system exhibits impurity-induced magnetic ordering at $T_{\rm N}=2.5$ K, at which magnetic susceptibility exhibits a cusplike maximum in the temperature scan. With increasing pressure, $T_{\rm N}$ increases. This can be attributed to the increase in the effective exchange interaction between the unpaired spins due to the shrinkage of the triplet gap in intact dimers. With further increase in pressure, the ground state undergoes the crossover to the uniform antiferromagnetic phase due to the complete closing of the triplet gap. This crossover occurs at $P\simeq 1.3$ kbar. The impurity-induced and pressure-induced uniformly ordered phases are contiguous, and thus, they are the same phase.

\begin{acknowledgments}
The authors thank A. Oosawa and H. -J. Mikeska for stimulating discussions. 
This work was supported by a Grant-in-Aid for Scientific Research and the 21st Century COE Program ``Nanometer-Scale Quantum Physics'' at Tokyo Tech., both from the Ministry of Education, Culture, Sports, Science and Technology of Japan.
\end{acknowledgments}

\end{document}